\documentclass[12pt]{article}
\usepackage[dvips]{graphicx}
\usepackage[myheadings]{fullpage}
\usepackage{tikz}
\usepackage{amsmath}
\usepackage{color}

\usetikzlibrary{shapes.geometric, arrows}
\tikzstyle{basicfit} = [rectangle, rounded corners, minimum width=2cm, minimum height=1cm,text centered, draw=black, fill=red!60, text width=2cm]
\tikzstyle{select} = [rectangle, rounded corners, minimum width=2cm, minimum height=1cm,text centered, draw=black, fill=blue!60, text width=2cm]
\tikzstyle{unfit} = [rectangle, rounded corners, minimum width=2cm, minimum height=1cm,text centered, text= black!60, draw=black!60, text width=2cm]
\tikzstyle{fit} = [rectangle, rounded corners, minimum width=2cm, minimum height=1cm,text centered, draw=black!60, text width=2cm]
\tikzstyle{arrow} = [thick,->,>=stealth]
\tikzstyle{nonarrow} = [very thin,->,>=stealth, draw= black!60]

\title{A Latent Trait Model for Multivariate Longitudinal Data With Two Sources of Measurement Error}

\author{Amy E. Nussbaum \thanks{Corresponding Author: anussbaum@smu.edu}
	\and Cornelis J. Potgieter 
	\and Department of Statistical Science \\ Southern Methodist University, Dallas, TX 75205
\and Michael Chmielewski
\and Department of Psychology\\ Southern Methodist University, Dallas, TX 75205}

\date{}

\begin{document}

\maketitle
\begin{abstract}
	Personality traits are latent variables, and as such, are impossible to
measure without the use of an assessment. Responses on the assessments can
be influenced by both transient (state-related) error and measurement error, obscuring the true trait levels. Typically, these assessments utilize Likert scales, which yield only discrete data. The loss of information due to the discrete nature of the data represents an additional challenge in assessing the ability of these instruments to measure the latent trait of interest.

This paper is concerned with parameter estimation in a model relating a latent variable, as well transient error and measurement error components when data are longitudinal and measured using a Likert scale. Two methods for parameter estimation are detailed: correlation reconstruction, a method that uses polychoric
correlations, and maximum likelihood implemented using a Stochastic EM algorithm. These methods are applied to a motivating dataset of 440
college students taking the Big Five inventory twice in a two month period.
\end{abstract}

\section{Introduction}

A common approach in latent variable modeling is the assumption that the
latent variables follow some type of continuous distribution. Specifically,
analysis often proceeds under the assumption of normality (see Fleeson \cite%
{Fleeson01}, Breckler \cite{Breckler90}, and Finney and DiStefano \cite%
{Finney06} for examples and details). However, data collected on individual
assessment items are often measured using a Likert scale, and as such, is
inherently discrete. For instance, subjects responding to a statement using
a five point Likert scale reply with a number from one to five representing
how strongly they agree or disagree with the statement. The use of such
scales represents data coarsening (a loss of information). To accurately estimate model parameters, the discrete nature of the data must be properly taken into account.

Parameter estimation in latent variable models based on polytomous outcome data is not a new topic, with groundbreaking work being done by Rasch \cite{Rasch61} and Andrich \cite{Andrich78}. An overview of the topic and references to recent literature can be found in \cite{Bartholomew01}. The contribution of this paper lies in allowing for two independent sources of error, called transient error and measurement error, that obscure the latent trait of interest. Identifiability of both sources of error necessitates data being longitudinal in nature. 

Two methods of parameter estimation are considered. The first of these, herein called correlation reconstruction, is a method that relies on polychoric
correlation estimation, see the work of Olsson (\cite{Olsson79A} and \cite{Olsson79B}) for development. The second method is maximum likelihood, which is implemented using a variation of the EM algorithm. The EM algorithm, developed by Dempster, Laird, \& Rubin \cite{Dempster77}, is a method for implementing maximum likelihood in incomplete data problems with latent variable models being only one common application. In this paper, a Stochastic EM algorithm \cite{Nielsen00} is implemented. This is a specific application of the more general Monte Carlo EM algorithm, see \cite{Wei90} and \cite{Levine01}, which is useful when conditional expectation terms occurring in the EM log-likelihood function are approximated using Monte Carlo integration. 

The present work was motivated by data collected from 440 students from the
University of Iowa taken at two different time points two weeks apart, see  Chmielewski and Watson \cite{Chmielewski09}. These
students took the Big Five Inventory (BFI), a widely used personality
assessment consisting of forty-four statements on five separate
traits: Agreeableness, Conscientiousness, Extraversion, Neuroticism, and
Openness. Subjects respond to the statements using a five point Likert scale. The BFI is a widely used measure of the Big Five with well established
psychometric properties (\cite{John99}).

It is useful to contextualize instruments such as the BFI using latent
state-trait models, developed by Rolf Steyer and Manfred Schmitt in the
nineties. Steyer and Schmitt \cite{Steyer90} define a trait as ``an enduring
property of a person, at least with respect to the time span considered."
These traits are not expected to change over a short period of time.
However, person's state can obscure the true level of his or her trait.
States are less consistent than traits. They can change depending on both
internal and external circumstances, such as ``one's present state of mind,
mood, self-consciousness, or on specific situational influences" \cite%
{Steyer90}. If the goal of a personality assessment is to measure trait
levels, states represent a systematic type of measurement error, otherwise
known as transient error or state error.

The stability of trait measurements can have ``implications for the
diagnosis of clinical disorders, [...] help determine the utility of
therapeutic interventions, and influence decisions regarding whether
individuals should be rehabilitated or placed into long-term supervision" 
\cite{Chmielewski09}. Because of the serious consequences of trait
measurement, it is vital that the instruments built to measure traits are
doing so effectively. Neglecting potential sources of error can lead to
severe repercussions, and it may be that many psychologists do not consider
the ramifications of their decisions. Chmielewski and Watson write,
``Indeed, it seems that the potential influence of error is often ignored
and that the general tendency is for all observed change to be considered
indicative of true change in the underlying construct" \cite{Chmielewski09}.
They point out that if the presence of transient error is acknowledged in
analysis of personality assessments, traits are actually more stable over
time than previously believed. Furthermore, they provide several examples of
analyses of data corrected for transient error and show that using a
correction can lead to different conclusions than those made without the
correction.

The paper proceeds as follows. In the next section, a latent variable model with transient error and measurement error components is proposed. Subsequent sections outline the correlation reconstruction and Stochastic EM algorithm approaches used for parameter estimation. Thereafter, a simulation study is done to compare the performance of these methods. After presenting an application of the methodology to one of the scales of the motivating data, conclusions are presented.

\section{Latent Variable Model Structure}

Let $\mathbf{Y}_{i} =\{Y_{ijt}\}_{\: j=1,\ldots,J, \: t=1,\ldots,T}$ denote the
Likert-scale response of individual $i$ to the $J$ items at $T$ timepoints.
The observed data are assumed to represent a coarsened version of the true continuous latent response to each item. Specifically,
let $\mathbf{X}_{i} =\{X_{ijt}\}_{\: j=1,\ldots,J, \: t=1,\ldots,T}$ denote the
latent response vector of individual $i$. For individual $i$, the latent
trait is denoted $Z_{i}$ with transient error denoted $e_{it}$ and
measurement error denoted $\varepsilon_{ijt}$. It is assumed that $Z_{i}$, $%
e_{it}$ and $\varepsilon_{ijt}$ are standard normal random variables. The
latent response $X_{ijt}$ is assumed to be of the form 
\begin{equation}  \label{model0}
X_{ijt}=\mu_j + \sigma_j Z_i + \tau_j e_{it} +\gamma_j \varepsilon_{ijt}.
\end{equation}
Here, $\sigma_{j}$ represents the strength with which item $j$ measures the true latent trait, while $\tau_{j}$ and $\gamma_{j}$ represent the sizes of the transient error and measurement error associated with item $j$. Note that the transient error is time-specific, but does not vary across items at a given period in time. On the other hand, measurement error is both item-specific and
time-specific.

Next, the relationship between the true latent vector $\mathbf{X}_{i}$ and
the observed response vector $\mathbf{Y}_{i}$ is formalized. One common model is to assume that Likert-scale observations are obtained from a $K$ parameter
cut point model, which is equivalent to the Likert scale consisting of $K+1$ categories. Formally, the cut points of item $j$ are denoted $\mathbf{z}_{j}=(z_{j1},z_{j2}, ... z_{jK})$ and the observed data are related to the latent response data thought the function $d(X_{ijt}| \mathbf{z}_{j} )$
where 
\begin{equation}  \label{truncation_model}
Y_{ijt} = d(X_{ijt} | \mathbf{z}_{j}) = \left\{ 
\begin{array}{ll}
1 & \text{if } X_{ijt} < z_{j1} \\ 
2 & \text{if } z_{j1} \leq X_{ijt} < z_{j2} \\ 
\vdots &  \\ 
K + 1 & \text{if } z_{jK} \leq X_{ijt}.%
\end{array}
\right.
\end{equation}

This model offers a great deal of flexibility, but requires the estimation
of a large number of cut points for any given scale. Future research will explore ways of reducing the number of cut points being estimated while trying to retain flexibility.

Next, note that the data coarsening resulting from the transformation (\ref{truncation_model}) means that the means and variances associated with latent variables $X_{ijt}$ are no longer identifiable from the Likert-scale
data. It is therefore assumed without loss of generality that $\mu_{j}=0$
and $\sigma^2_j + \tau^2_j + \gamma^2_j = 1$ for $j=1,\ldots,J$. The model in (\ref{model0}) now becomes 
\begin{equation}  \label{modelrewrite}
X_{ijt}=\sigma_j Z_i + \tau_j e_{it} +\gamma_j \varepsilon_{ijt}.
\end{equation}
Under this simplified model, $\sigma_j^2$ represents the proportion of total item variance due to signal strength, and $\tau_j^2$ and $\gamma_j^2$ represent the proportions of total item variance due to transient error and measurement error. Furthermore, the covariance matrix and correlation matrix for the model are identical. The correlation structure derived from model (\ref{modelrewrite}) has a block-patterned structure. Let $\boldsymbol{\Sigma}$ denote the covariance matrix of $\mathbf{X}$, then

\begin{center}
\begin{equation}  \label{sigmamat}
\boldsymbol{\Sigma} = \left[ 
\begin{array}{cccc}
\mathbf{A} & \mathbf{B} & \cdots & \mathbf{B} \\ 
\mathbf{B} & \mathbf{A} & \cdots & \mathbf{B} \\ 
\vdots & \vdots & \ddots & \vdots \\ 
\mathbf{B} & \mathbf{B} & \cdots & \mathbf{A}%
\end{array}
\right]
\end{equation}
\end{center}

where

\begin{center}
$\mathbf{A} = \left[ 
\begin{array}{cccc}
1 & \sigma_1 \sigma_2 + \tau_1 \tau_2 & \cdots & \sigma_1 \sigma_J + \tau_1
\tau_J \\ 
\sigma_2 \sigma_1 + \tau_1 \tau_2 & 1 & \cdots & \sigma_2 \sigma_J + \tau_2
\tau_J \\ 
\vdots & \vdots & \ddots & \vdots \\ 
\sigma_J \sigma_1 + \tau_J \tau_1 & \sigma_J \sigma_2 + \tau_J \tau_2 & 
\cdots & 1%
\end{array}
\right]$
\end{center}

and

\begin{center}
$\mathbf{B} = \left[ 
\begin{array}{cccc}
\sigma_1^2 & \sigma_1 \sigma_2 & \cdots & \sigma_1 \sigma_J \\ 
\sigma_2 \sigma_1 & \sigma_2^2 & \cdots & \sigma_2 \sigma_J \\ 
\vdots & \vdots & \ddots & \vdots \\ 
\sigma_J \sigma_1 & \sigma_J \sigma_2 & \cdots & \sigma_J^2%
\end{array}
\right]_. $
\end{center}

\bigskip

It seems prudent to emphasize that the block-patterned matrix $\boldsymbol{\Sigma}$ corresponds to the correlation structure of the latent vector $\mathbf{X}$ and not the Likert-scale vector $\mathbf{Y}$. The correlation structure of $\mathbf{Y}$ can be derived in a straight-forward manner, but expressions are double summations of terms involving the bivariate normal cdf. Furthermore, these are neither useful for the purposes of parameter estimation, nor for model interpretation, and are not included here.

\section{The Correlation Reconstruction Approach}

In this section, a moment-based approach to estimating the item-specific cut points is described. Thereafter, the general idea behind correlation reconstruction, which makes use of polychoric correlation estimation, is outlined.

\subsection{Cut Point Estimation}

From model (\ref{modelrewrite}), each latent variable $X_{ijt}$ has a standard normal distribution. Therefore, $P(Y_{ijt}\leq k)=P(X_{ijt} \leq z_{jk})=\Phi (z_{jk})$ from which
it follows that $z_{jk}=\Phi^{-1}\lbrack P(Y_{ijt}\leq k) \rbrack$. Here, $%
\Phi$ and $\Phi^{-1}$ denote the standard normal cdf and quantile functions.
This motivates cut point estimators 
\begin{equation}  \label{z_hat_part1}
\hat{z}_{jkt} = \Phi^{-1} \left( \frac{ \sum_{i = 1}^n \mathcal{I} \left(
Y_{ijt} \leq k \right) + 1}{n + 2} \right)
\end{equation}
where $\mathcal{I}$ denotes the indicator function. The proposed estimator,
hereafter referred to as the inverse normal method of moments estimator, includes a small-sample adjustment (adding one in the numerator and two in the
denominator) to avoid the possibility of estimated cut points equal to $ \pm \infty $. Extensive numerical work suggests that this adjustment has a two-fold effect: while it does result in a slight increase in the bias of the cut point estimators, there is an overall reduction in RMSE. Furthermore, the
bias introduced is by this adjustment is asymptotically negligible. Recognizing that cut points are not functions of time, data across multiple time points can be combined to improve estimation, namely 
\begin{equation}  \label{z_hat_part2}
\hat{z}_{jk} = \frac{1}{T} \sum_{t = 1}^T \hat{z}_{jkt}.
\end{equation}

\subsection{Polychoric Correlation}

Below is a brief outline of how polychoric correlation estimation can be implemented to recover the correlation between two latent variables when only their Likert-scale counterparts are observed. Polychoric correlation was first implemented by $\boldmath{XXX}$. Let $(X_1,X_2)$ follow a bivariate normal distribution with zero mean, unit variance and correlation coefficient $\rho$. Let $(Y_1,Y_2)$ denote the associated Likert-scale vector, $Y_1 = d(X_1 | \mathbf{z}_1)
$ and $Y_2 = d(X_2 | \mathbf{z}_2)$ where $\mathbf{z}_{j} = \{z_{j1}, z_{j2},
z_{j3}, z_{j4}\}$ for $j=1,2$. For each pair of values $(y_1, y_2)$, $y_1, y_2 \in \{1,...,K+1\}$, define $p_{y_1,y_2}=P(Y_1 = y_1, Y_2 = y_2)$. The empirical estimate of $p_{y_1,y_2}$ based on data $\left(Y_{1i},Y_{2i} \right) $, $i=1,\ldots,n$,  is given by
\begin{equation*}
\hat{p}_{y_1,y_2} = \frac{\sum_{i = 1}^{n} \mathcal{I} (Y_{1i} = y_1,Y_{2i} = y_2) }{n}.
\end{equation*}
The associated theoretical probability is given by 
\begin{align*}
p_{y_1,y_2} = & \; P(z_{1,y_1-1} \leq X_1 \leq z_{1,y_1}, z_{2,y_2-1} \leq
X_2 \leq z_{2,y_2}) \\
= & \; \Phi(z_{1,y_1}, z_{2,y_2} | \rho) - \Phi(z_{1,y_1}, z_{2,y_2-1} |
\rho) - \Phi(z_{1,y_1-1}, z_{2,y_2} | \rho) + \Phi(z_{1,y_1-1}, z_{2,y_2-1}
| \rho)
\end{align*}
where $\Phi(\cdot,\cdot|\rho)$ is the bivariate normal cdf with standard normal marginals and correlation coefficient $\rho$. The likelihood function, expressed in terms of these empirical and
theoretical probabilities, is given by 
\begin{equation}  \label{CR_likelihood}
\mathcal{L} = \prod^{K+1}_{y_1=1} \prod^{K+1}_{y_2=1} \{ P(Y_1 = y_1, Y_2 =
y_2) \}^{n p_{y_1,y_2}}.
\end{equation}

The above likelihood function can be maximized simultaneously over $(\mathbf{%
z}_1,\mathbf{z}_2,\rho)$, but this requires maximizing a nonlinear function
in $2K+1$ dimensions. Alternatively, a marginal likelihood function for $\rho
$ can be constructed by substituting the inverse normal method of moment
estimators $\mathbf{\hat{z}}_1$ and $\mathbf{\hat{z}}_2$, with components calculated as in (\ref{z_hat_part2}), into the likelihood function.
As discussed by Olsson, (\cite{Olsson79A}), the differences with regards to
efficiency between the full likelihood and marginal plug-in likelihood are very small, and the use of a marginal likelihood function has the advantage of reduced computational labor. In this paper, the marginal plug-in approach will be used.

\subsection{Correlation Reconstruction}

The outlined procedure in the previous subsection for calculating the polychoric correlation between two Likert-type items can
be generalized to estimating the correlation structure for all $J$ items on
the assessment. Recall that $\boldsymbol{\Sigma}$ denotes the covariance
matrix for latent vector $\mathbf{X}$. First, all cut points for all items are estimated by applying the inverse normal method according to (\ref{z_hat_part1}) and (\ref{z_hat_part2}). 
Let $\tilde{\rho}_{(j,s),(k,t)}$ denote the correlation estimate
reconstructed from observed pairs $(Y_{ijs}, Y_{ikt})$, $%
i=1,\ldots,n$. Define $(\tilde{\mathbf{A}}_t)_{jk}=\tilde{\rho}_{(j,t),(k,t)}
$ whenever $j\not=k$ and $(\tilde{\mathbf{A}}_t)_{jj}=1$ otherwise. Also define 
$(\tilde{\mathbf{B}}_{st})_{jk}=\tilde{\rho}_{(j,s),(k,t)}$ where $s\not=t$.
The matrix of so-called reconstructed correlation coefficients, $\boldsymbol{\tilde{\Sigma}}$, is given by

\begin{center}
\begin{equation}  \label{sigmamat_tilde}
\boldsymbol{\tilde{\Sigma}} = \left[ 
\begin{array}{cccc}
\mathbf{A}_1 & \mathbf{B}_{12} & \cdots & \mathbf{B}_{1T} \\ 
\mathbf{B}_{21} & \mathbf{A}_{2} & \cdots & \mathbf{B}_{2T} \\ 
\vdots & \vdots & \ddots & \vdots \\ 
\mathbf{B}_{T1} & \mathbf{B}_{T2} & \cdots & \mathbf{A}_{T}%
\end{array}
\right].
\end{equation}
\end{center}

While the reconstructed correlation matrix (\ref{sigmamat_tilde}) is
asymptotically unbiased for the true correlation structure, it does not
typically conform to the block-diagonal structure as described in
(\ref{sigmamat}). Therefore, attempting to use $\boldsymbol{\tilde{\Sigma}}$ directly to estimate the $3J$ parameters $(\boldsymbol{\sigma},\boldsymbol{\tau},\boldsymbol{\gamma})$ results in an over-identified system of equations. One way to circumvent this problem is to use a matrix norm such as the Frobenius norm to minimize the distance between estimate $\boldsymbol{\tilde{\Sigma}}$ and parameterized matrix $\boldsymbol{\Sigma}$. Specifically, define the Frobenius norm of an $m \times
n$ matrix $\mathbf{D}$, 
\begin{equation*}
||\mathbf{D}||_F = \sqrt{\sum_{i = 1}^m \sum_{j = 1}^n d_{ij}^2}
\end{equation*}
where $d_{ij}$ is the element in the $i^{th}$ row and $j^{th}$ column of
matrix $\mathbf{D}$. Then, the minimum distance estimators of $(\boldsymbol{\sigma},\boldsymbol{\tau},\boldsymbol{\gamma})$ are obtained by minimizing
\begin{equation*}
H(\boldsymbol{\sigma},\boldsymbol{\tau},\boldsymbol{\gamma}) = ||\boldsymbol{\tilde{\Sigma}}-\boldsymbol{\Sigma}||_F.
\end{equation*}
Note that $\gamma_j = \sqrt{1-\sigma_j^2-\tau_j^2}$ and one therefore only needs to minimize this function in terms of $\sigma_j$
and $\tau_j$, $j = 1, ..., J$.

In this paper, the choice of the Frobenius norm is motivated by moderate
robustness considerations. As the Frobenius norm is the square root of the
Euclidean distance between the two matrices, it is more robust against
outliers than, say, the Euclidean distance. However, there may be considerations other than robustness which motivate the practitioner and those could result in the use of a different matrix norm. 

\section{Maximum Likelihood using a Stochastic EM Algorithm}

Direct maximization of the likelihood for the model being considered is numerically difficult, as the latent trait component common to all items at all time points and the transient error component common to all items at a given time point result in likelihood function involving multiple
integrals with no closed-form solutions. Difficulties with similar models have been addressed by \cite{Nielsen00}. However, the EM algorithm is a method for
finding maximum likelihood estimators when analyzing data sets with missing
values and/or latent responses and is ideal for use in the present setting.
First developed by Dempster, Laird, and Rubin \cite{Dempster77}, the EM
algorithm starts with the complete data log-likelihood assuming the latent
variables were observed. The algorithm then iterates between an expectation
step (E-step) where the expectation of the complete data log-likelihood function is evaluated conditional on the observed data and using an estimate of the model parameters. Next, in the maximization step (M-step), the function
obtained in the E-step is maximized to update the parameter estimates. These
two steps are alternated until convergence is reached.

To be more precise, let $\ell (\theta | \mathbf{X},\mathbf{Y})$ denote the complete data log-likelihood and let $\theta^{(r)}$ denote the value of the
parameter estimates after $r$ iterations of the algorithm. During the E-step
of the algorithm, the function $Q(\theta \big| \theta^{(r)}) =
E_{\theta^{(r)}}[\ell (\theta | \mathbf{X},\mathbf{Y}) \big|\mathbf{Y}]$ is evaluated and then during the M-step, $Q(\theta \big| \theta^{(r)})$ is
maximized in terms of $\theta$ in order to obtain new parameter
estimates $\theta^{(r+1)}$. This is repeated until convergence of the
parameter vector. In the present setting, the parameter vector is given by $%
\theta = (\boldsymbol{\sigma},\boldsymbol{\tau},\mathbf{z})$ where $%
\boldsymbol{\sigma} = ( \sigma_1, ... \sigma_J), \boldsymbol{\tau} = (
\tau_1, ... \tau_J)$, and $\mathbf{z} = ( \mathbf{z}_1, ... \mathbf{z}_J)$
with $\mathbf{z}_j = (z_{j1}, ... z_{jK})$. The cut points $z_{jk}$ are
defined as before, but for convenience the notation $z_{j0}=-\infty
$ and $z_{j,K+1}=\infty$ for $j=1,\ldots,J$ is also introduced.

Recall that the latent vector $\mathbf{X}$ follows a
multivariate normal distribution with density function 
\begin{equation}  \label{eq:density}
f_{\mathbf{X}}\left( \mathbf{x} \right) = \left( (2 \pi)^{-\textstyle{\frac{%
JT}{2}}} |\boldsymbol{\Sigma}|^{-\textstyle{{\frac{1}{2}}}} \right) \exp
\left( -\frac{1}{2} \boldsymbol{x}^{\prime }\boldsymbol{\Sigma}^{-1} 
\boldsymbol{x} \right)
\end{equation}
where $\boldsymbol{x}=\left( x_{11}, \ldots, x_{JT} \right)$ and $%
\boldsymbol{\Sigma}$ is the block-patterned covariance matrix in (\ref%
{sigmamat}). The responses $Y_{ij}$ conditional upon the latent variables $X_{ij}$ are independent of one another with 
\begin{equation}  \label{eq:y_given_x}
f_{Y_{jt}|X_{jt}}\left( y_{jt}|x_{jt} \right) = \mathcal{I}
\left(z_{j,y_{jt}-1} \leq x_{jt} < z_{j,y_{jt}} \right),\quad
j=1,\ldots,J,\,t=1,\ldots,T,\ .
\end{equation}
The joint distribution of response vector $\mathbf{Y}$ and latent
vector $\mathbf{X}$, found by combining (\ref{eq:density}) and (\ref%
{eq:y_given_x}), is 
\begin{align*}
f_{\mathbf{X}, \mathbf{Y}}\left(\mathbf{x}, \mathbf{y} \right) = & \left( (2
\pi)^{-\textstyle{\frac{JT}{2}}} |\boldsymbol{\Sigma}|^{-\textstyle{\frac{1}{%
2}}} \right) \exp \left( -\frac{1}{2} \boldsymbol{x}^{\prime }\boldsymbol{%
\Sigma}^{-1} \boldsymbol{x} \right) \\
& \times \prod_{j = 1}^J \prod_{t = 1}^T \mathcal{I} \left(z_{j,y_{jt}-1}%
\leq x_{jt}<z_{j,y_{jt}}\right).
\end{align*}
for $x_{jt}\in \mathrm{I\!R}$ and $y_{jt} \in \left\{1,\ldots,K\right\}$ for
all $j,t$. Notice from the above joint distribution of $(\mathbf{X},\mathbf{Y%
})$ that the distribution of $\mathbf{X}|\mathbf{Y}$ is a multivariate
truncated normal.

Now, given individual response vectors $\mathbf{y}_{i}=\left(
y_{i11},\cdots, y_{iJT}\right)$ and associated unobserved latent response
vectors $\mathbf{x}_{i}=\left( x_{i11},\cdots, x_{iJT}\right)$ for $%
i=1,\ldots, n$, the complete-data likelihood is given by 
\begin{equation*}
L\left( \theta \right) = \prod_{i=1}^{n} \left( f_{\mathbf{X}_{i}} \left( 
\mathbf{x}_{i} \right) \times \prod_{j = 1}^J \prod_{t = 1}^T \mathcal{I}
\left(z_{j,y_{ijt}-1}\leq x_{ijt}<z_{j,y_{ijt}}\right) \right).
\end{equation*}
The function $Q(\theta \big| \theta^{(r)})$ (excluding constants not involving the parameters) is
\begin{align*}
Q ( \theta | \hat{\theta}^{(r)} )  = -\frac{n}{2} \log|\boldsymbol{\Sigma}| -\frac{1}{2} \sum_{i=1}^{n} E_{%
\hat{\theta}^{(r)}}\left[ \mathbf{X}_{i}^{\prime }\boldsymbol{\Sigma}^{-1} 
\mathbf{X}_{i} |\mathbf{Y}_{i} = \mathbf{y}_{i} \right] \\
+ \sum_{i = 1}^n \sum_{j = 1}^J \sum_{t = 1}^T E_{\hat{\theta}^{(r)}} %
\left[ \log \mathcal{I} \left(z_{j,Y_{ijt}-1} \leq X_{ijt} < z_{j,Y_{ijt}}
\right) |Y_{ijt} = y_{ijt}\right].
\end{align*}
This function is separable in that it is the sum of a function
depending only on $\boldsymbol{\sigma}$ and $\boldsymbol{\tau}$, the
parameters used to define elements of the covariance matrix $\boldsymbol{%
\Sigma}$, and a function depending only on the cut points $\mathbf{z}_{j}$, $j=1,\ldots,J$. Therefore, define 
\begin{equation}  \label{eq:c1}
Q_1 ( \theta | \hat{\theta}^{(r)} ) = -\frac{n}{2} \log|%
\boldsymbol{\Sigma}| -\frac{1}{2} \sum_{i=1}^{n} E_{\hat{\theta}^{(r)}}\left[
\mathbf{X}_{i}^{\prime }\boldsymbol{\Sigma}^{-1} \mathbf{X}_{i} |\mathbf{Y}%
_{i} = \mathbf{y}_{i} \right]
\end{equation}
and 
\begin{equation}  \label{eq:c2}
Q_2 ( \theta | \hat{\theta}^{(r)} ) = \sum_{i = 1}^n \sum_{j =
1}^J \sum_{t = 1}^T E_{\hat{\theta}^{(r)}} \left[ \log \mathcal{I}
\left(z_{j,Y_{ijt}-1} \leq X_{ijt} < z_{j,Y_{ijt}} \right) |Y_{ijt} = y_{ijt}%
\right].
\end{equation}

\bigskip

Notice that the conditional expectations in (\ref{eq:c1}) and (\ref{eq:c2}%
) are equivalent to evaluating expectations of functions of multivariate
truncated normal random variables. Recently, there has been renewed interest
in explicit evaluation of the joint moments of this distribution, for
example, see \cite{Wilhelm10} and \cite{Horrace05} (references to previous
work can be found therein). However, neither the quadratic form in (\ref%
{eq:c1}) nor the log-indicator function in (\ref{eq:c2}) have been
considered in the literature. One approach would be to evaluate these using numerical integration techniques, but in the present paper we choose to follow a
different approach. 

The Monte Carlo EM (MCEM) algorithm (\cite{Wei90}, \cite{Nielsen00}) is a
modification of the EM algorithm used when direct evaluation of the
conditional expectations is cumbersome, but it is possible to sample from
the conditional distribution. In the present setting, this requires sampling from a multivariate truncated normal distribution. Algorithms for doing so have been considered by \cite{RodriguezYam04}%
, \cite{Wilhelm10}, and \cite{Botev15}. The MCEM algorithm replaces the
conditional expectations of the form $E_{\hat{\theta}^{(r)}}\left(h(\mathbf{X},\mathbf{Y}) | 
\mathbf{Y}=\mathbf{y} \right)$ by approximations
\begin{equation*}
\hat E_{\hat{\theta}^{(r)}}\left(h(\mathbf{X},\mathbf{Y}) | 
\mathbf{Y}=\mathbf{y} \right) =
\frac{1}{m} \sum_{m=1}^{M} h(\tilde{\mathbf{X}}^{(m)},\mathbf{y})
\end{equation*}
where $\tilde{\mathbf{X}}^{(m)}$, $m=1,\ldots,M$ are sampled from the
distribution $[\mathbf{X}|\mathbf{Y}=\mathbf{y}]$ assuming $\theta=\hat{\theta}^{(r)}$.
This paper implements a variation of the MCEM algorithm known as the Stochastic EM (StEM) algorithm. The StEM algorithm is an implementation of MCEM
that sets $M=1$, i.e. at each iteration of the algorithm, only one sample point is drawn per observation. This algorithm is discussed in greater detail in \cite{Nielsen00}.
Formally, the StEM equations analogous to (\ref{eq:c1}) and (\ref{eq:c2}) are
given by 
\begin{equation}  \label{eq:c1v2}
\hat Q_1 ( \theta | \hat{\theta}^{(r)} ) = -\frac{n}{2} \log|%
\boldsymbol{\Sigma}| -\frac{1}{2} \sum_{i=1}^{n} \tilde{\mathbf{X}}%
_{i}^{\prime }\boldsymbol{\Sigma}^{-1} \tilde{\mathbf{X}}_{i}
\end{equation}
and 
\begin{equation}  \label{eq:c2v2}
\hat Q_2 ( \theta | \hat{\theta}^{(r)} ) = \sum_{i = 1}^n \sum_{j
= 1}^J \sum_{t = 1}^T \log \mathcal{I} \left(z_{j,y_{ijt}-1} \leq \tilde{X}%
_{ijt} < z_{j,y_{ijt}} \right).
\end{equation}
where $\tilde{\mathbf{X}}_{i}$ was drawn from the conditional distribution $[\mathbf{X}|\mathbf{Y}=\mathbf{y}_{i}]$ with parameter values $\theta=\hat{\theta}^{(r)}$.

Equation (\ref{eq:c1v2}) is continuous in the parameters $\boldsymbol{\sigma}
$ and $\boldsymbol{\tau}$ and therefore standard optimization methods can be used. In this paper, the Barzilai-Borwein method as implemented in the R package BB was used to maximize (\ref{eq:c1v2}), see \cite{Varadhan09} for details. Special care needs to be taken when maximizing (\ref{eq:c2v2}). In fact, this function does not have a unique maximum and most standard numerical approaches will give
erratic results. However, it is possible to define a closed-form estimator of
the cut points. Define sets $\mathcal{S}_{j,k}=\left\{\left(i,t%
\right):Y_{ijt}=k\right\}$, $k=1,\ldots,K$. Inspection of (\ref{eq:c2v2})
reveals that it is constant as a function of $z_{jk}$ for all for all $%
z_{jk} \in \left[ \max_{(i,t) \in \mathcal{S}_{j,k}} \tilde{X}_{ijt},
\min_{(i,t) \in \mathcal{S}_{j,k+1}} \tilde{X}_{ijt} \right]$. Any value
inside this interval would be a valid maximum likelihood estimator of $z_{jk}
$. The non-uniqueness of the maximum is sidestepped by defining

\begin{equation*}
\hat{z}_{jk}^{(r)} = \frac{\max_{(i,t) \in \mathcal{S}_{j,k}} \tilde{X}%
_{ijt} + \min_{(i,t) \in \mathcal{S}_{j,k+1}} \tilde{X}_{ijt}}{2}
\label{zhat}
\end{equation*}
for the $r^{th}$ step of the StEM algorithm and for $j = 1, \ldots, J$ and $%
k = 1, \ldots, K-1$. Note that this is one possible maximum likelihood estimator.

In implementation, the simulation and maximization steps of the StEM
algorithm are repeated $R$ times and the final parameter estimates are
defined to be the averages of the estimators obtained at each of the $R$
steps. Specifically, 
\begin{align*}
\hat{\sigma}_j & = \frac{1}{R} \sum_{r=1}^R \hat{\sigma}_j^{(r)} \\
\hat{\tau}_j & = \frac{1}{R} \sum_{r=1}^R \hat{\tau}_j ^{(r)} \\
\hat{z}_{jk} & = \dfrac{1}{R} \sum_{r=1}^R \hat{z}_{jk}^{(r)}
\end{align*}
for $j = 1, \ldots, J$ and $k = 1, \ldots, K$. While it is preferable to use
larger values of $R$, this also requires more resources in terms of time and
computation. The effect of choosing $R$ will be investigated in a Monte
Carlo simulation study in the next section.

\section{Simulation Study}

Several simulation studies were performed to assess and compare the
performance of correlation reconstruction and
StEM-based maximum likelihood methods. For implementation of the StEM
algorithm, samples from the multivariate truncated normal distribution were
generated using rejection sampling as implemented in \cite{Botev15}.

In the simulation study, latent responses were simulated according to model (\ref{modelrewrite}) using specified parameter values $(\boldsymbol{\sigma},\boldsymbol{\tau},\boldsymbol{\gamma})$.
The simulated latent responses were then converted to Likert-scale
responses according to (\ref{truncation_model}) using specified cut points $\mathbf{z}_{j}$ for the $j^{th}$ item, $j=1,\ldots,J$.

For the simulation presented here, data were generated from a model with $J = 5$
items and $T=2$ time points. The model parameters were $\boldsymbol{%
\sigma} = (0.8^{1/2}, 0.7^{1/2}, 0.6^{1/2}, 0.5^{1/2}, 0.4^{1/2})$, $\boldsymbol{\tau} = (0.1^{1/2}, -0.15^{1/2},-0.2^{1/2}, 0.25^{1/2}, 0.3^{1/2})$, and $\boldsymbol{\gamma} = (0.1^{1/2}, 0.15^{1/2}, 0.2^{1/2}, 0.25^{1/2}, 0.3^{1/2})$. These were chosen to have items ranging from large signal strength (80\% of variability) to small signal strength (40\% of variability). Samples of size $n=100$ and $n=250$ were simulated from this latent response model and then converted to Likert scales with $K=5$ using cut points $\mathbf{z}_{j}=(-1.2, -0.5, 0.4, 0.8)$ for 
$j=1, 5$ and $\mathbf{z}_{j}=(-0.85, -0.25, 0.25, 0.85)$ for $j= 2, 3, 4$.

 A total of $M=1000$ data sets were generated in this way. For each, both correlation reconstruction and the StEM algorithm were implemented. For the StEM algorithm, $R = 1000$ iterations were used to also study the effect that the number of iterations has on the final parameter estimates. The StEM algorithm was also implemented with $R=5000$ iterations, but results did not differ substantially from the case $R=1000$ and are therefore not included.

\begin{figure}[h]
	\label{IndexPlots}\centerline{\includegraphics[keepaspectratio=T, scale = 0.55]{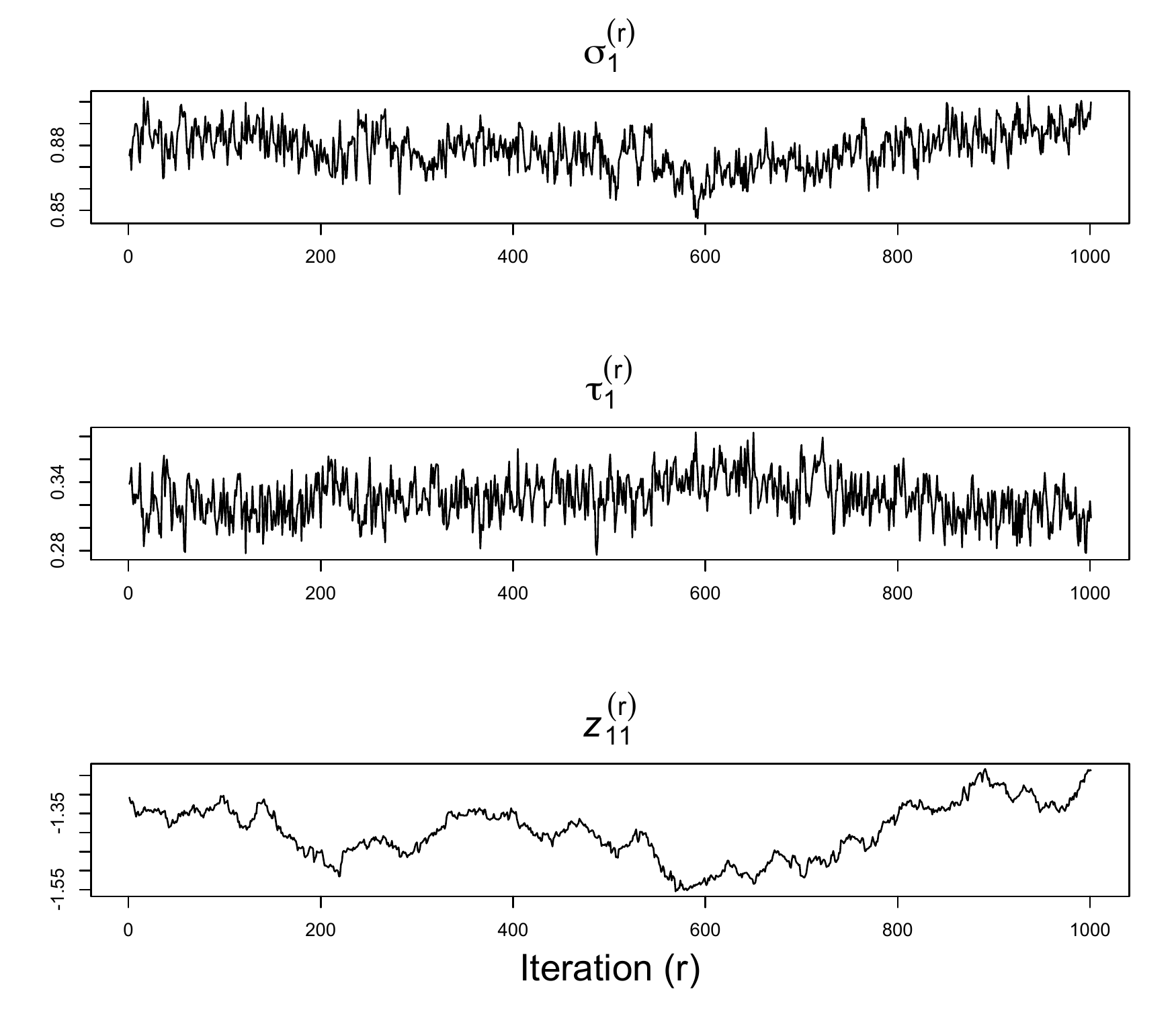}} 
	\caption{Index Plots for Stochastic EM Algorithm}
\end{figure}

\begin{figure}[!ht]
	\label{acfplots}\centerline{\includegraphics[keepaspectratio=T, scale = 0.55]{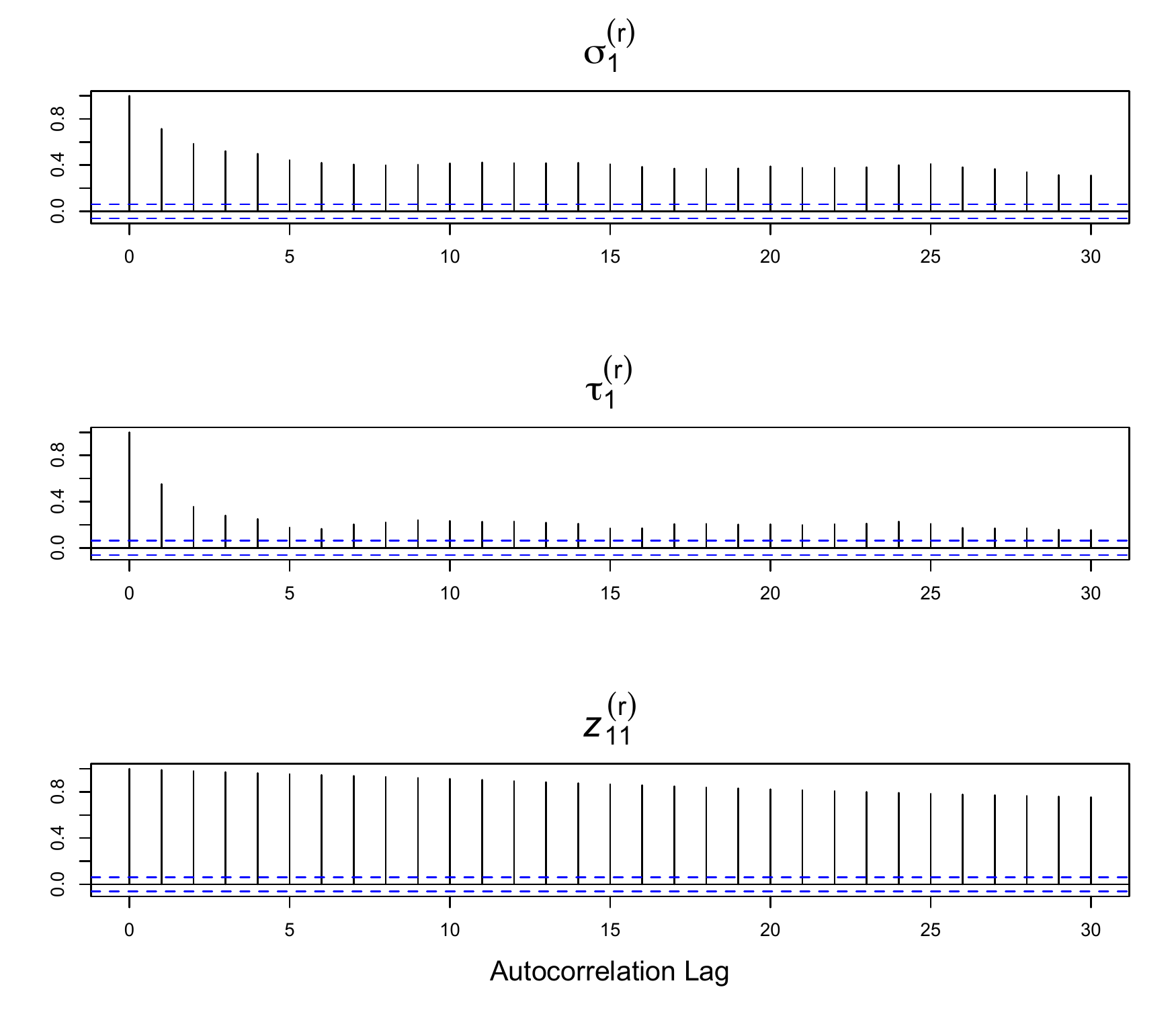}} 
	\caption{Autocorrelation Plots for Stochastic EM Algorithm}
\end{figure}

To illustrate the StEM algorithm, index plots for parameter estimates $\hat{\sigma}_1^{(r)},\hat{\tau}_1^{(r)},\hat{z}_{11}^{(r)})$, $r=1,\ldots,1000$ corresponding to a simulated data set are shown in Figure \ref{IndexPlots}. Index plots for other parameter estimates have similar appearance and are excluded. Autocorrelation plots for these same parameter estimates are shown in Figure \ref{acfplots}.

The index plots for $\hat{\sigma}_1^{(r)}$ and $\hat{\tau}_1^{(r)}$ show good mixing and the corresponding autocorrelation plots show that the autocorrelation quickly decreases. This indicates that fewer iterations of the StEM algorithm are required to estimate the parameters of the covariance matrix. On the other hand, both the index plot and autocorrelation plot for $\hat{z}_{11}^{(r)}$ show persistent autocorrelation even at large lags. This indicates that the algorithm may need to run for a large number of iterations to accurately estimate cut points.
\bigskip
\begin{center}
	Figures \ref{IndexPlots} and \ref{acfplots} About Here
\end{center}
\bigskip
Tables \ref{FullRhoRMSEJ5T2} and \ref{FullCutsRMSEJ5T2} summarize the simulation results using RMSE as a measure of performance.
\bigskip
\begin{center}
	Tables \ref{FullRhoRMSEJ5T2} and \ref{FullCutsRMSEJ5T2} About Here
\end{center}
\bigskip
Consider Table \ref{FullRhoRMSEJ5T2}. It is clear that the RMSE of the estimators $\hat{\sigma}_j$ increases as  the value of $\sigma_j$ decreases ($\sigma_1>\ldots>\sigma_5$). This holds true both for correlation reconstruction and StEM. On the other hand, the RMSE of the estimators $\hat{\tau}_j$ increases as the (absolute) value of $\tau_j$ increases ($|\tau_1|<\ldots<|\tau_5|$). The StEM estimators, which are approximate maximum likelihood estimators, tend to have smaller RMSE than the correlation reconstruction estimators. This effect is especially pronounced at sample size $n=100$ where, for example, the RMSE of $\hat{\sigma}_1$ decreases by over $36\%$ when $n=100$, but decreases by less than $1\%$ when $n=250$. 

The results in Table \ref{FullCutsRMSEJ5T2} are noteworthy in that the inverse normal method of moments (MM) cut point estimators consistently have smaller RMSE than the StEM algorithm with $R = 1000$. Again, this trend persisted with $R=5000$. As with estimation of the covariance parameters, the difference in performance of the two methods is only pronounced at the smaller sample size $n=100$.

The superiority of the MM approach for cut point estimation as compared to StEM can easily be explained. Firstly, the MM cut point estimators defined in (\ref{z_hat_part1}) incorporate an intentional bias - adding one in the numerator and two in the denominator before applying the inverse normal transform. As noted previously, this leads to a reduction in RMSE in small samples. Secondly, the StEM algorithm is, per definition, stochastic in nature. This leads to an added source of variability when computing these estimators. This source of variability decreases as $R$ increases, but because of persistent autocorrelation present in $\hat{z}^{(r)}_{\cdot \cdot}$ at large lags, a very large value of $R$ may be required before this source of variablity becomes negligible.

Although the StEM algorithm results in estimators theoretically optimal in an RMSE sense, it comes with a high cost. The average time to calculate estimates for a sample size of $250$ individuals taking an assessment with $J = 5$ items and $T = 2$ time points is slightly over $60$ minutes while correlation reconstruction takes around two minutes on the same personal computer. As a general observation based on extensive simulation results, correlation reconstruction performs nearly as well as StEM when $T>2$ or when the sample size becomes large. Based on the considerable computational cost associated with the StEM algorithm, correlation reconstruction is recommended as a convenient alternative that performs nearly as well even in moderate sample sizes.

\section{Analysis of Motivating Data}

The motivating data for the techniques developed in this paper comes from Chmielewski and Watson (2009). The data was collected from 440 students at
the University of Iowa who completed the Big Five Inventory (BFI) twice with assessment opportunities two weeks apart. The
inventory consists of forty-four statements on five separate traits:
Agreeableness, Conscientiousness, Extraversion, Neuroticism, and Openness.
The scales have between eight and ten items associated with them. For the
sake of brevity, only estimates for the Openness scale are presented in the paper.

Adjectives associated with Openness include intelligent, imaginative, and
perceptive, and researchers have also linked Openness to differentiated
emotions, aesthetic sensitivity, and the need for variety to Openness.
Individuals with low Openness levels are typically described as favoring
conservative values, while those rated high tended to enjoy aesthetic
impressions and have wide interests \cite{McCrae87}. The Openness scale has
a total of ten items.

Estimates of the cut points calculated the inverse normal method of moments
and the stochastic EM algorithm are shown in tables \ref{ocuts} and \ref%
{stemocuts}. There differences in cut point estimates based on the two methods are negligible.
performance.
\bigskip
\begin{center}
	Tables \ref{ocuts} and \ref{stemocuts} About Here
\end{center}
\bigskip
Estimates of the signal strength, transient error, and measurement error components are shown in Table \ref{rcopars} (correlation reconstruction) and Table \ref{stemopars} (StEM). Note that estimated parameters are reported on a squared scale, as this scale has the natural interpretation of being a proportion of variance attributed to, respectively, the latent trait, transient error and measurement error. Also, as the transient error coefficients are allowed to be negative, a superscript $(-)$ will indicate that the parameter estimate on the original scale was negative. Item number corresponds to numbering on the BFI questionnaire.
\bigskip
\begin{center}
	Tables \ref{rcopars} and \ref{stemopars} About Here
\end{center}
\bigskip
Firstly, note that the two estimation methods give very similar results for the measurement error components $\hat{\gamma}_j^2$. On many of the items, close to $50\%$ of the latent item variance can be attributed to measurement error. Item 35 is notable in that around $99\%$ of the latent item variance can be attributed to measurement error. Of course, it is impossible to say whether this is only specific to the target population of this study, or if this item gives poor results in general. 

When considering the latent trait components $\hat{\sigma}_j^2$, correlation reconstruction and StEM show a slight difference. Consistently, the signal strength estimate under correlation reconstruction is larger than that under StEM. The converse is true for the transient error components $\hat{\tau}_j^2$. However, while there are differences, the parameter estimates under both methods of estimation convey similar information: it would seem that item 44 is the single best measure of the latent trait with upward of $65\%$ of variability explained by the latent trait. Item 30 is a close second, while other items are mainly in the $25\%$ to $30\%$ range. As noted previously, item 35 does not seem to measure the latent trait in the present setting.

Also of note is the estimation of transient error components $\hat{\tau}_j^2$. The work done here was specifically to develop methodology for capturing transient error. It seems this is an important factor in the model. While several items have estimated transient error effects below $5\%$, items 5, 15, 25 and 40 all have sizeable transient error effects present.

Some may question the interpretation of the transient error component estimates that have negative signs. Conditional on an individual's trait level, the sign of $\tau_j$ indicates which direction the responses tend to vary in relation to one another. For instance, item 5 of the BFI reads, ``I am someone who is original, comes up with new ideas" whereas item 41 reads, ``I am someone who prefers work that is routine." If transient error caused a higher response for a particular individual to item 5, the response to item 41 would likely be lower for the same individual.

\section{Conclusion}

In this paper, two methods for estimating parameters in longitudinal Likert-scale study with both transient error and measurement error were developed. The first approach relies on polychoric correlation estimation and then uses a matrix norm to estimate the parameters of interest. On the other hand, a stochastic EM algorithm is presented as a method of approximating the maximum likelihood estimators. While the StEM method has superior performance as measured by RMSE, implementation is computationally intensive and time-consuming. In the simulation study performed, correlation reconstruction performs nearly as well as the StEM algorithm for samples of size $n=250$. 

The techniques developed have been applied to a motivating data set. In doing
so, evidence of transient error in a realistic setting was identified and
the efficacy of the techniques has been demonstrated. Questions for future research include latent trait prediction at the individual level after model estimation has been performed, as well as model selection when deciding whether or not to include transient error and measurement error components.

\vspace{\fill}\pagebreak



\pagebreak

\section{Tables and Figures}

\begin{table}[h]
\begin{center}
\begin{tabular}{|c||c|c||c|c|}
\hline
& \multicolumn{2}{|c||}{$n = 100$} & \multicolumn{2}{|c|}{$n = 250$} \\ 
\hline
Estimator & \; \; \; \; CR \; \; \; \; & StEM & \; \; \; \; CR
\; \; \; \; & StEM \\ \hline
$\hat{\sigma}_1$ & 0.036 & 0.023 & 0.015 & 0.014 \\ \hline
$\hat{\sigma}_2$ & 0.040 & 0.031 & 0.020 & 0.019 \\ \hline
$\hat{\sigma}_3$ & 0.046 & 0.040 & 0.025 & 0.025 \\ \hline
$\hat{\sigma}_4$ & 0.053 & 0.048 & 0.030 & 0.031 \\ \hline
$\hat{\sigma}_5$ & 0.059 & 0.057 & 0.034 & 0.035 \\ \hline\hline
$\hat{\tau}_1$ & 0.049 & 0.048 & 0.030 & 0.030 \\ \hline
$\hat{\tau}_2$ & 0.056 & 0.054 & 0.034 & 0.033 \\ \hline
$\hat{\tau}_3$ & 0.060 & 0.058 & 0.038 & 0.036 \\ \hline
$\hat{\tau}_4$ & 0.058 & 0.056 & 0.037 & 0.036 \\ \hline
$\hat{\tau}_5$ & 0.063 & 0.060 & 0.038 & 0.036 \\ \hline
\end{tabular}%
\end{center}
\caption{RMSE for correlation reconstruction (CR) and StEM ($R=1000$) estimators of $\protect\sigma_j$ and $\protect\tau_j$, $j=1,\ldots,J$,
for simulation with $J = 5$, $T = 2$.}
\label{FullRhoRMSEJ5T2}
\end{table}

\begin{table}[tbp]
\begin{center}
\begin{tabular}{|c||c|c||c|c|}
\hline
& \multicolumn{2}{|c||}{$n = 100$} & \multicolumn{2}{|c|}{$n = 250$} \\ 
\hline
True Value & \; \; \; \; CR. \; \; \; \; & StEM & \; \; \; \; CR
\; \; \; \; & StEM \\ \hline
$\hat{z}_{11}$ & 0.138 & 0.155 & 0.094 & 0.099 \\ \hline
$\hat{z}_{12}$ & 0.114 & 0.120 & 0.076 & 0.077 \\ \hline
$\hat{z}_{13}$ & 0.113 & 0.121 & 0.075 & 0.076 \\ \hline
$\hat{z}_{14}$ & 0.122 & 0.134 & 0.079 & 0.082 \\ \hline\hline
$\hat{z}_{21}$ & 0.122 & 0.135 & 0.079 & 0.082 \\ \hline
$\hat{z}_{22}$ & 0.111 & 0.116 & 0.072 & 0.073 \\ \hline
$\hat{z}_{23}$ & 0.109 & 0.115 & 0.071 & 0.073 \\ \hline
$\hat{z}_{24}$ & 0.115 & 0.127 & 0.076 & 0.081 \\ \hline\hline
$\hat{z}_{31}$ & 0.113 & 0.124 & 0.075 & 0.079 \\ \hline
$\hat{z}_{32}$ & 0.103 & 0.108 & 0.070 & 0.071 \\ \hline
$\hat{z}_{33}$ & 0.104 & 0.109 & 0.070 & 0.071 \\ \hline
$\hat{z}_{34}$ & 0.116 & 0.128 & 0.076 & 0.080 \\ \hline\hline
$\hat{z}_{41}$ & 0.114 & 0.126 & 0.073 & 0.077 \\ \hline
$\hat{z}_{42}$ & 0.102 & 0.110 & 0.065 & 0.068 \\ \hline
$\hat{z}_{43}$ & 0.098 & 0.105 & 0.065 & 0.067 \\ \hline
$\hat{z}_{44}$ & 0.111 & 0.123 & 0.070 & 0.074 \\ \hline\hline
$\hat{z}_{51}$ & 0.124 & 0.137 & 0.079 & 0.086 \\ \hline
$\hat{z}_{52}$ & 0.102 & 0.107 & 0.067 & 0.070 \\ \hline
$\hat{z}_{53}$ & 0.101 & 0.109 & 0.067 & 0.070 \\ \hline
$\hat{z}_{54}$ & 0.108 & 0.122 & 0.070 & 0.075 \\ \hline
\end{tabular}%
\end{center}
\caption{RMSE for correlation reconstruction (CR) and StEM ($R=1000$) estimators of cut point estimators for simulation with $J = 5$, $T = 2$.}
\label{FullCutsRMSEJ5T2}
\end{table}

\begin{table}[tbp]
\begin{center}
\begin{tabular}{|c||c|c|c|c|c|c|c|c|c|c|c|}
\hline
\multicolumn{11}{|c|}{Openness} &  \\ \cline{1-11}
Cuts & 5 & 10 & 15 & 20 & 25 & 30 & 35 & 40 & 41 & 44 &  \\ \cline{1-11}
$\hat{z}_1$ & -1.75 & -2.08 & -1.82 & -1.69 & -1.76 & -1.56 & -1.31 & -1.93
& -1.49 & -1.17 &  \\ \cline{1-11}
$\hat{z}_2$ & -1.00 & -1.43 & -0.80 & -0.87 & -0.64 & -0.80 & -0.21 & -0.90
& -0.53 & -0.38 &  \\ \cline{1-11}
$\hat{z}_3$ & -0.14 & -0.71 & 0.06 & -0.14 & 0.30 & -0.17 & 0.55 & -0.04 & 
-0.02 & 0.22 &  \\ \cline{1-11}
$\hat{z}_4$ & 1.14 & 0.74 & 1.11 & 0.99 & 1.44 & 0.69 & 1.52 & 1.10 & 0.82 & 
1.01 &  \\ \cline{1-11}
\end{tabular}%
\end{center}
\caption{Inverse Normal Method of Moments Cut Point Estimates Openness}
\label{ocuts}
\end{table}
\begin{table}[tbp]
\begin{center}
\begin{tabular}{|c||c|c|c|c|c|c|c|c|c|c|c|}
\hline
\multicolumn{11}{|c|}{Openness} &  \\ \cline{1-11}
Cuts & 5 & 10 & 15 & 20 & 25 & 30 & 35 & 40 & 41 & 44 &  \\ \cline{1-11}
$\hat{z}_1$ & -1.75 & -2.11 & -1.80 & -1.70 & -1.75 & -1.54 & -1.33 & -1.98
& -1.45 & -1.15 &  \\ \cline{1-11}
$\hat{z}_2$ & -0.97 & -1.43 & -0.78 & -0.86 & -0.61 & -0.77 & -0.23 & -0.86
& -0.52 & -0.37 &  \\ \cline{1-11}
$\hat{z}_3$ & -0.12 & -0.70 & 0.08 & -0.14 & 0.31 & -0.14 & 0.55 & -0.03 & 
-0.01 & 0.22 &  \\ \cline{1-11}
$\hat{z}_4$ & 1.17 & 0.75 & 1.14 & 0.98 & 1.50 & 0.71 & 1.54 & 1.12 & 0.85 & 
1.06 &  \\ \cline{1-11}
\end{tabular}%
\end{center}
\caption{stEM Cut Point Estimates for Openness}
\label{stemocuts}
\end{table}

\begin{table}[tbp]
\begin{center}
\begin{tabular}{|c||c|c|c|c|c|c|c|c|c|c|c|}
\hline
\multicolumn{11}{|c|}{Openness} &  \\ \cline{1-11}
Item ($j$)  & 5 & 10 & 15 & 20 & 25 & 30 & 35 & 40 & 41 & 44 &  \\ \cline{1-11}
$\hat{\sigma}_j^2$ & 0.41 & 0.28 & 0.31 & 0.35 & 0.39 & 0.66 & 
0.02 & 0.40 & 0.40 & 0.65 &  \\ \cline{1-11}
$\hat{\tau}_j^2$ & 0.15 & 0.06 & 0.18 & 0.08 & 0.14 & 0.02$^{(-)}$ & 0.00$^{(-)}$ & 0.16 & 0.13$^{(-)}$ & 0.04$^{(-)}$ &  \\ \cline{1-11}
$\hat{\gamma}_j^2$ & 0.43 & 0.66 & 0.51 & 0.57 & 0.47 & 0.33 & 
0.99 & 0.44 & 0.47 & 0.31 &  \\ \cline{1-11}
\end{tabular}%
\end{center}
\caption{Correlation Reconstruction Covariance Parameter Estimates Openness}
\label{rcopars}
\end{table}
\begin{table}[tbp]
\begin{center}
\begin{tabular}{|c||c|c|c|c|c|c|c|c|c|c|c|}
\hline
\multicolumn{11}{|c|}{Openness} &  \\ \cline{1-11}
Item ($j$) & 5 & 10 & 15 & 20 & 25 & 30 & 35 & 40 & 41 & 44 &  \\ \cline{1-11}
$\hat{\sigma}_j^2$ & 0.33 & 0.25 & 0.24 & 0.27 & 0.29 & 0.69 & 
0.01 & 0.31 & 0.44 & 0.74 &  \\ \cline{1-11}
$\hat{\tau}_j^2$ & 0.24 & 0.12 & 0.24 & 0.13 & 0.22 & 0.00 & 0.00
& 0.30 & 0.04$^{(-)}$ & 0.02$^{(-)}$ &  \\ \cline{1-11}
$\hat{\gamma}_j^2$ & 0.43 & 0.63 & 0.52 & 0.59 & 0.49 & 0.31 & 
0.99 & 0.39 & 0.52 & 0.24 &  \\ \cline{1-11}
\end{tabular}%
\end{center}
\caption{StEM Covariance Parameter Estimates Openness}
\label{stemopars}
\end{table}

\end{document}